# Optimizing performance in Basketball: A Game-Theoretic Approach to Shot Percentage Distribution in a team

Aditya Singh[1][0009-0001-5431-6460]

[1] George Washington University, Washington D.C. 20052, USA
Email: adityasingh@gwu.edu

**Abstract.** In this paper we propose a shot percentage distribution strategy among the players of a basketball team to maximize the score that can be achieved by them. The approach is based on the concepts of game theory that are related to network flow. The paper starts with drawing similarity between network flow problem and passing sequence in basketball. The concept of Price of anarchy was applied in the basketball. Different strategies that can be used by teams are evaluated and compared with the proposed strategy that consider the players shooting behavior as the game progresses. The work also looks at the interaction of different participating players and how their collective behavior can be used to achieve an optimum performance. The paper explains that giving the ball to the best player of the team to take a shot might not be the best strategy to maximize the team's overall score.

## 1   Introduction

Basketball is a popular sport played by two teams with five players each. The team roster can have a maximum of twelve players with unlimited substitution allowed. The end goal is to score more points than the other team by throwing the ball through a hoop mounted on a ten foot high backboard at both ends of the court.

Among the two teams, at any given moment the ball is with one team hence they are on offense and the other team is on defense. The ball is moved between players of the attacking team, a single instance of ball movement is called a pass. At the end of this complex movement of ball, a player takes an attempt to shoot a basket. We can see this movement of ball as a network flow problem as we will see in the next section.

Network traffic flow is a well researched domain and game theory is one of the tools that can be used to optimize the network. In subsequent sections we will draw similarities between network flow problem and basketball, go through the key game theory concepts, and understand the mathematics behind the optimizing process. The paper will conclude with results, limitations and future scope of this work.

## 2 Literature Review

### 2.1 Network flow problem

There are many interesting optimization problems in the domain of transportation, fluids and several other domains that can be viewed as a network flow problem. The general theory of networks tries to solve these problems in many diverse contexts using mathematical tools like but not limited to combinatorics and linear algebra [2].

Every network requires two entities: *nodes* and *edges* (also called arcs, link). The edges of a network can be unidirectional or bidirectional. Sometimes there can be a value associated to an edge that can be called its "capacity". A network is often showed pictorially as in Figure 1.

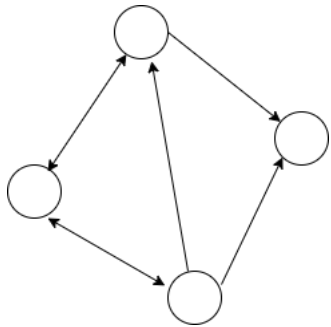

**Fig. 1.** Generic network with edges and nodes

### 2.2 Game theory on networks

Game theory is a branch of mathematics that helps in understanding how individuals, groups, organization or countries interact with each other in a system where the decision taken by one entity can impact the choices available to the other entities in that system. Generally, the participating entities are called *"players"* in game theory, we will be using them interchangeably. Every option the entity of network chooses results in an outcome or reward which is called *"payoffs"*.

Game theory can be applied to understand a system of network where multiple entities are competing or cooperating to optimize the flow of information or resources through the network. "Price of Anarchy" is a concept in game theory that calculates the degradation of network due to selfish behavior of a player who seek of optimize their payoff without taking into consideration the global collective optimum. Game theory provides a very powerful and effective framework to understand the dependencies between the entities in the network, identify the equilibrium and to design solutions to incentivize cooperative behavior between players of the system.

**Price of Anarchy in traffic flow network – Braess Paradox**

In traffic network flow, price of anarchy [6] can be clearly visualized often termed as "Braess Paradox". The following example is taken from the work of Skinner, B. (2009) [1].

Consider a simple network with two nodes A and B (represents cities here). There are two edges between them, edge 1 is a highway and edge 2 is a small sub lane along the highway. The network can be represented pictorially as shown in Figure 2.

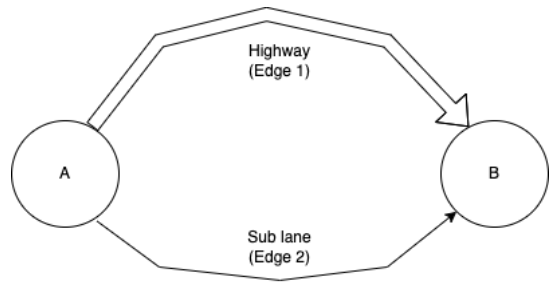

**Fig. 2.** Two nodes A and B connected by two edges termed as Highway and sub lane

Assuming that there are 10 cars, the time it takes to go from point A to point B using highway is a constant 10 units. If using sub lane, then the time linearly depends on the number of cars present in the sub lane i.e., if there is just 1 car then 1 minute, if there are 3 cars then 3 minutes and so on. The equilibrium of system can be achieved when every car takes the sub lane, as from a duration point of view there is no incentive in using the highways. This equilibrium is called "Nash Equilibrium" where every player (i.e., car) is acting selfishly without thinking about the entire system consisting of other players. It can be proved mathematically that the best course of action would be that 5 cars take the highway and other 5 cars take the sub lane. The complete mathematical solution can be found in the work of Skinner, B. (2009) [1].

The difference between the overall system payoff in Nash equilibrium and the payoff when everyone is acting towards the optimum of entire system is called "Price of Anarchy".

### 2.3 Similarity between basketball and network flow

Basketball if viewed from a distance is a network flow problem. Each player is a node, every pass is a possible *edge* (or arc) between the players. If a player takes the final shot, then the basket mounted on the backboard is a *sink* (or end node). There are infinite options of sequences which a team can opt for in an offensive position. The number would be much huge if we consider the defensive actions that is taken by the opposition. In this scope of this work, we will only consider the offensive action in the game of basketball. In a network form, a generic gameplay in basketball can be represented as shown in figure 3 Every directional edge from a node to another is a pass completed.

If we try to draw similarity between the traffic network and basketball then every passing sequence starts with a player let say node A, there are some intermediate nodes and edges between them. Finally, a player takes the shot to score a basket, basket is like node B in traffic network discussed in the section above.

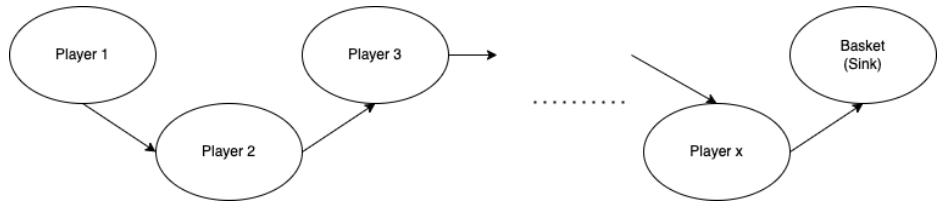

**Fig. 3.** Passes between players is represented by directional arrows, final player who takes the shot is represented by 'x' and there is a node for basket which is like sink node in a flow network.

It might feel like the best possible passing sequence would be the one that gives the highest possible chance of success. It might be a sequence where the player with the best shooting accuracy takes the final shot, but we will show later that such choice of passing sequence is equivalent to "Nash equilibrium" or selfish and it doesn't guarantee the best efficiency from a team point of view.

### 2.4 Previous Research

There is very limited research done in the field of optimizing gameplay in basketball. Most of the research is performed to find the best sequence of passing or to find the trajectory of shooting to maximize total points without considering the changes in players shooting capability once the game starts. Moreover, nearly all the work focuses on using machine learning (see Javadpour [3] et. al.) that tries to achieve the result that is close to Nash equilibrium which might not be the most optimal result in the real world scenario. There are very few research pieces when it comes to using game theory to optimize the performance of the team.

Work done by Brian Skinner (2009) [1] talks about using game theory to optimize team's performance in basketball. This research work perfectly explains the use of game theory and explains how game theory can be used to achieve the true global optimum for a basketball team.

### 2.5 Research gaps

As discussed in the previous section, most of the research around optimizing team's performance is in machine learning domain. The only drawback about the Skinner, B [1] work is that it just talks about how it's not always a good option to pass the ball to the player with best shooting accuracy in a game. The paper talks about shooting behavior of just one player with all other players from the team having a constant shooting behavior. The research broadly focuses on single player in a multi player system.

There is a need to explore domains like game theory to optimize team performance in basketball. There is a need to extend the work of Skinner, B. to a multi-player system where every player has an independent shooting behavior. It is important to understand how different players with different shooting behaviors can collaborate in a team to maximize the team's utility, to understand different strategies and to compare the payoffs each of those strategies can deliver. This paper will try to cover all these gaps discussed above.

## 3    Methodology

### 3.1    Terminologies

Before diving into the details, it is important to understand the terminologies that will be used in this work. Most of these are basketball terminologies that will be used in the later sections. There are few terms that are not generally used in basketball but are required to understand the later sections.

1. Free throw attempts (FTA): When a player gets an opportunity to score from a free throw line without any interference from the opposition. These are awarded because of foul committed by an opposition player.
2. Field goal attempts (FGA): It refers to any attempt taken to try to score points by shooting basketball in the opponent's basket. Field goal can be attempted from various distances, and they are the fundamental way to score points in the game.
3. Points score (PS): This is the total points scored by a player or the team. In this paper it will always be the points scored by a player unless stated otherwise.
4. Total shots (TS): This is the summation of FTA and FGA of a player.
5. True Shooting percentage (TS%): This is an advanced statistics and it best thought of as a field goal percentage adjusted for free throws and field goal shots (see Kubatko et. al., 2007 [4]). TS% is defined by the formula given in equation 1.

$$TS\% = \frac{(0.5*(points\ scored))}{((field\ goal\ attempts) + 0.44*(free\ throw\ attempts))} \qquad (1)$$

6. Fraction of team shots (FTS): To calculate the fraction or % of team shots taken (equation 2) by each player per game, we used the total time the player played for in a game, total shots the player took and the total team shots. This was a calculated for each game.

$$FTS = \left(\frac{players\ shots/game}{team\ shots/game}\right) * \left(\frac{48\ minutes/game}{player\ minutes/game}\right) \tag{2}$$

These are all the technical terms that will be used in the later sections.

### 3.2 Dataset

Data was collected for Washington Wizards which is an NBA team based out of Washington DC. The data about overall players performance, teams' performance and individual players statistics on a game by game basis was collected for the 2022-23 season. The data was collected from *https://www.basketball-reference.com/* which accumulates the data of teams and players.

Each players' data had 30 features associated for each match day. After careful analysis we reduced the number of required features to five. Those five features are minutes played, games started, field goal attempts (FGA), free throw attempts (FTA) and points scored. We further calculated TS% and FTS for each player on a game to game basis.

### 3.3 Players shooting behavior

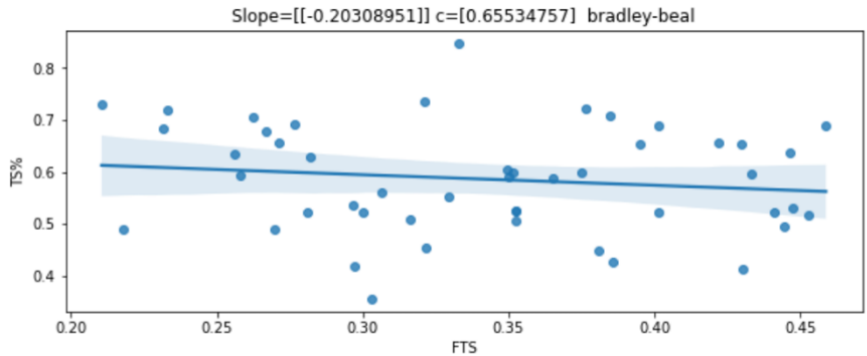

**Fig. 4.** The graph above shows the inverse relationship between TS% and FTS. The slope and y-intercept of the linear relationship is mentioned in the graph title along with the player's name. The slope and intercept are stored for every player, and it defines the players shooting behavior i.e. shot accuracy

Calculating the player's shooting behavior is a very difficult task and it is impacted by a lot of factors. First and foremost, the defensive action performed by the opposing team plays a significant factor in deciding a player's shooting behavior. It can also be impacted by factors that are beyond the scope of game. In the book "Basketball on Paper (2004)" [5], author Dean Oliver envisioned that there is an inverse relationship between TS% and FTS of a player. This relationship between TS% and FTS remains a highly theoretical concept but this inverse relationship is the closest imitation to the real world scenario.

Using the TS% and FTS calculated earlier, we find a linear function that inversely relates the two factors for every player of Washington Wizards. The linear relationship between TS% and FTS for a Bradley Beal is shown in figure 4.

### 3.4 Optimizing function and constraints

Let us assume a player A takes x percent of total team shots, using the behavior model created in the above section we can find the actual TS% for a given percentage of shots as shown in equation 3 (it is very much like sub lane, one of the paths in the traffic network example discussed in section 2.2).

$$(TS\% \; for \; x\% \; shots) = \left(x_{player\,A} slope_{player\,A} + intercept_{player\,A}\right) \quad (3)$$

With the calculated TS% for x% shots, we can find the actual shot % that would result in a score when a player takes x% of team shots. This value will be called utility. The sum of utility contributed by all five players of the team will be the team payoff. We can define the TS% of a player as a function of percentage of shots taken (i.e. $f_{player\,A}(x)$). To calculate the utility for $player_A$ we can use equation (3):

$$(Utility)_{player\,A} = x_A \; f_{player\,A}(x) \quad (4)$$

In a team of five players, with each having their own shooting behavior we can create the objective function F that is to be maximized as

$$F = x_1 f_1(x_1) + x_2 f_2(x_2) + x_3 f_3(x_3) + x_4 f_4(x_4) + x_5 f_5(x_5) \quad (5)$$

In the equation 5, F is the objective function that we want to maximize, $x_i$ is the percent of team shots $player_i$ takes in a game, and $f_i$ is the shooting behavior function that we had calculated earlier. There are some constraints over the variables used in the objective function that is to be considered. The sum of shot % taken by every player should be equal to 1 (equation 6). The shot % of any player should not exceed 40% of overall team shots in a game (equation 7). This value was determined by past behavior observed across teams in different seasons. The utility contributed by each player should never be negative (equation 8).

$$x_1 + x_2 + x_3 + x_4 + x_5 = 1 \tag{6}$$

$$0.0 \leq x_i \leq 0.40 \tag{7}$$

$$x_i f_i(x_i) \geq 0 \tag{8}$$

We can maximize the objective function by staying in the bounds of constraint. The value of the objective function is what the team can achieve by taking into consideration the player's shooting behavior.

## 4 Results

We ran the simulation model for every combination of five players who are part of the team and have played at least a defined number of games. There were two groups created for the players. Group 1 had players who started at least 30 games (regular starters) and group 2 had all the players which were part of the team roster and have played (different from starting the game) a certain number of games. For each of those groups we evaluated four different scenarios in which a team can operate.

We looked at four scenarios to compare the payoff a team (group of five players) can get. The most obvious scenario, where the player with the best shooting ability (the player with highest intercept value) takes the shot for most of the times. Then we moved to scenario where every player tries to contribute an equal utility value to the final objective function, we also looked at scenario where every team member is taking an equal percentage of shots i.e. 20% of total team shots each. Finally, we concluded with the proposed model we defined above, which considers the constraints, and shooting behavior model which we had calculated earlier.

### 4.1 Group of regular starters

There were only 7 players who started more than 30 games, an observation which aligns with what we have seen in earlier season. On evaluating the result for all four different scenarios in which a team can operate for every combination of 5 players. Our proposed strategy model was better by 2% than the second best approach. The second best approach being that the player with best intercept value takes the most percentage of shots which is not a very ideal scenario in the real world but can be stated as the best case scenario. The proposed strategy model was better by 9% than the strategy where every player takes an equal percentage of shots which is much closer to real world scenario. The comparison of different strategies can be seen in figure 5.

### 4.2 Group of all players on the team roster

There are a total of 14 players which were present on the team roster which satisfied our initial filtering of players based on various factors like minutes played and have played at least 10 games (not necessarily started the games). There were a total of 2002 different combination of players.

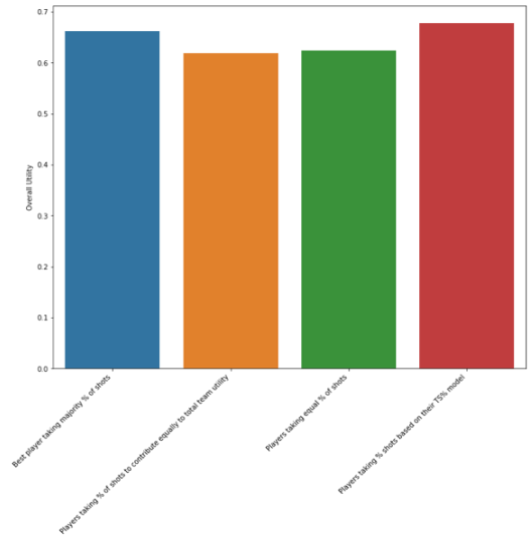

**Fig. 5.** Payoffs for different strategies for the group of regular starters. The proposed strategy which considers player's shooting behavior is shown in red. The utility value for proposed model is 0.678

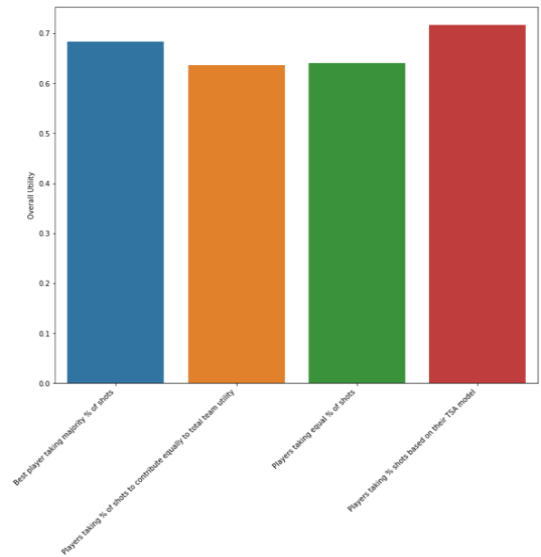

**Fig. 6.** Payoffs for different strategies for the group of all the players in the roster. The utility value for proposed model in this case is 0.717 (shown in red)

Our proposed model was better by 4% than the second best approach where the best player takes the most percentage of team shots. The result was much more prominent when it comes to scenarios that imitate the real world situation. Our strategy was 12% better than the strategy where every player takes an equal percentage of shots. The comparison of different strategies and our proposed strategy is shown in figure 6. The result obtained for this group was much more prominent.

## 5 Conclusion

In the first part we looked at the existing research that has been done in the area. We looked at the potential avenues and gaps in the work which has been done at the intersection of sports and game theory. The second part of the work looked at the methodology of the work starting with introducing the concepts that would be required to understand the work, then we deep dived into the data that has been used in the work. Later, we put our focus on modelling the player's shooting behavior using terms derived from player stats using linear regression and we derived the optimization function and walked through the constraints applied over the function. In the final section we looked the result and compared proposed strategy with other different strategies.

There are a lot of external factors that affects the performance of the team in basketball. This work doesn't take into consideration the defensive action that the opposition takes against the team with the ball. Defensive action is surely an important factor that affects the performance of the team. We only looked at the final arc of a passing sequence i.e. when the player takes a final shot that may or may not result in the increase of the score.

This paper provides the foundation and introduces optimization techniques for applying game theory concepts to basketball. We showed similarity between network flow and basketball, hence extending the concept of "Price of anarchy" to the game of basketball to find optimum distribution of shot percentage among the players of the team to maximize the points in a game of basketball. Sports is in general a domain where there is a lot of external influence, this makes analysis of sports very difficult. This paper provides a new perspective to improve the performance of basketball teams by finding the optimum distribution of shot attempts in the team to maximize points scored, a small step towards a more accurate quantitative analysis.